\documentclass[aps,prd,twocolumn,showpacs,amsmath,amssymb,floatfix]{revtex4-1}
\usepackage{epsfig}
\usepackage{graphicx}
\usepackage{dcolumn}
\usepackage{bm}
\usepackage{ltablex,booktabs}
\usepackage{overpic}
\usepackage{subfigure}
\usepackage{float}
\usepackage{color}
\usepackage{amsmath}
\usepackage{mathcomp}
\usepackage{mathrsfs}
\usepackage{multirow}
\usepackage{rotating}
\usepackage{amssymb}
\usepackage{gensymb}
\usepackage{amsmath}
\usepackage{tabularx}
\usepackage{overpic}
\usepackage{colortbl}
\usepackage{textcomp}
\usepackage{tabu}
\usepackage{xcolor}
\usepackage{makecell}
\usepackage{subfigure}
\usepackage[separate-uncertainty=true]{siunitx}
\usepackage{dcolumn}
\usepackage[unicode]{hyperref}
\usepackage{enumitem}
\input{bes3sym}

\def\pipipi{\pip\pim\pip}
\def\swave{${\cal S}$ wave}
\def\swavehy{${\cal S}$-wave}

\newcommand{\putat}[3]{\begin{picture}(0,0)(0,0)\put(#1,#2){#3}\end{picture}}
\newcommand{\RNum}[1]{\uppercase\expandafter{\romannumeral #1\relax}}

\begin{document}
\normalsize
\parskip=5pt plus 1pt minus 1pt

\title{ \boldmath Amplitude analysis of $D_s^{+} \rightarrow \pi^{+} \pi^{-} \pi^{+}$}
\author{
\begin{small}
\begin{center}
M.~Ablikim$^{1}$, M.~N.~Achasov$^{10,b}$, P.~Adlarson$^{67}$, S. ~Ahmed$^{15}$, M.~Albrecht$^{4}$, R.~Aliberti$^{28}$, A.~Amoroso$^{66A,66C}$, M.~R.~An$^{32}$, Q.~An$^{63,49}$, X.~H.~Bai$^{57}$
% ...
\\
\vspace{0.2cm}
(BESIII Collaboration)\\
\vspace{0.2cm} {\it
$^{1}$ Institute of High Energy Physics, Beijing 100049, People's Republic of China\\
% ...
$^{4}$ Bochum Ruhr-University, D-44780 Bochum, Germany\\
% ...
$^{10}$ G.I. Budker Institute of Nuclear Physics SB RAS (BINP), Novosibirsk 630090, Russia\\
% ...
$^{15}$ Helmholtz Institute Mainz, Staudinger Weg 18, D-55099 Mainz, Germany\\
% ...
$^{28}$ Johannes Gutenberg University of Mainz, Johann-Joachim-Becher-Weg 45, D-55099 Mainz, Germany\\
% ...
$^{32}$ Liaoning Normal University, Dalian 116029, People's Republic of China\\
% ...
$^{49}$ State Key Laboratory of Particle Detection and Electronics, Beijing 100049, Hefei 230026, People's Republic of China\\
% ...
$^{57}$ University of Jinan, Jinan 250022, People's Republic of China\\
% ...
$^{63}$ University of Science and Technology of China, Hefei 230026, People's Republic of China\\
% ...
$^{66}$ University of Turin and INFN, (A)University of Turin, I-10125, Turin, Italy; (B)University of Eastern Piedmont, I-15121, Alessandria, Italy; (C)INFN, I-10125, Turin, Italy\\
$^{67}$ Uppsala University, Box 516, SE-75120 Uppsala, Sweden\\
% ...
$^{b}$ Also at the Novosibirsk State University, Novosibirsk, 630090, Russia\\
% ...
}\end{center}

\vspace{0.4cm}
\end{small}
}

\date{\today} 

\begin{abstract} Utilizing the data set corresponding to an integrated
luminosity of $3.19$ fb$^{-1}$ collected by the BESIII detector at a
center-of-mass energy of 4.178 GeV, we perform an amplitude analysis
of the $\Ds\to\pip\pim\pip$ decay. The sample contains 13,797
    candidates with a signal purity of $\sim$80\%.  
The amplitude and phase of the contributing $\pi\pi$ \swave\ 
are measured based on a quasi-model-independent approach, 
along with the amplitudes and phases of
 the ${\cal P}$ and ${\cal D}$
waves parametrized by Breit-Wigner models. 
    The fit fractions of
different intermediate decay channels are also reported.

\end{abstract}
\maketitle

\section{Introduction} \label{sec:introduction} 

The decay $\Ds\to\pipipi$ is interesting due to its dominant \swave\
and relatively large branching fraction~\cite{FOCUS,babarpaper,CLEO:2013bae,pdg}. This provides an
opportunity to study the structure of the $\pi\pi$ \swave\ below 2~GeV
and improve our understanding of light scalar mesons such as
$f_0(980)$ and $f_0(1370)$, whose exact natures remain a mystery and
are open to different interpretations~\cite{pdg}.  The $f_0(980)$ is
particularly interesting as it is produced via hadronization of an
$s\bar{s}$ quark-antiquark pair close to the $K\bar{K}$ threshold. Its
couplings to both $\pi\pi$ and $K\bar{K}$ final states can be studied
in decays such as $\Ds\to\pipipi$ and $\Ds\to\KpKm\pip$. 
The study of
the  $\pi\pi$ \swave\ in $\Ds\to\pipipi$ 
can also shed light on the production
mechanism of $f_0(980)$~\cite{PRD94.096002}.

Besides the \swavehy\ amplitude, amplitude analysis of $\Ds\to
\pipipi$ can also offer vital information on the branching fraction of $\Ds\to
\rho^0\pip$. 
As pointed out in Ref.~\cite{Cheng:2016ejf}, $\Ds\to\rho^0\pip$ 
is unique because it is the only observed $D\to VP$
decay that the difference, not the sum, of $W$-annihilation amplitudes for the production of pseudoscalar meson
($P$) and vector meson ($V$) is involved. 
Neither the magnitudes
nor strong phases of the $A_{P,V}$ amplitudes can be determined
without the knowledge of the $\Ds\to \rho^0\pip$ branching
fraction. 
Therefore, based on the fit fraction determined in the 
$\Ds\to
\pipipi$ amplitude analysis, 
${\cal B}(\Ds\to \rho^0\pip)$ is a crucial
experimental input in the global analysis of two-body $D\to VP$ decays
in Ref.~\cite{Cheng:2016ejf}.


Amplitude analyses of $\Ds\to\pipipi$ have been performed previously
by the E687~\cite{E687}, E791~\cite{E791}, FOCUS~\cite{FOCUS}, and
BABAR~\cite{babarpaper} experiments. BABAR also reported the first
quasi-model-independent partial wave analysis (QMIPWA) to model
the \swavehy\ amplitude on this channel using
a relatively large data sample of 13,179 $\Ds$ candidates with a signal purity of 80\%. In
this paper, based on a 3.19~\invfb data sample collected with the
Beijing Spectrometer (BESIII) in 2016 at a center-of-mass energy
($E_{\rm c.m.}$) of 4.178 GeV, we present an amplitude analysis of
$\Ds\to\pipipi$ also based on the QMIPWA approach, with a data sample
comparable to the one used by BABAR, and a similar purity. At this
energy, $\Ds$ mesons are produced predominantly through the processes
$\epem\to D_s^{*\pm} D_s^{\mp}$. A single $D_s^+$ (or $D_s^-$)
is reconstructed by its final state particles. This analysis uses
$\Ds\to\pipipi$ and its charge conjugate channel. If one event 
contains more than one $D_s^\pm$
candidate, all candidates 
will be considered for further analysis. 
Charge-conjugate states are implied
throughout this paper.

The paper is organized as follows. Section~\ref{sec:dect} introduces
the BESIII detector and the data and Monte Carlo (MC) simulated
samples used in this analysis. Section~\ref{chap:event_selection}
gives an overview of the event selection technique and criteria.  The
details of the $\Ds\to\pipipi$ amplitude analysis are described in
Sec.~\ref{sec:ampana}, and the fit results are shown in
Sec.~\ref{fitting}. The systematic uncertainties on our measurements
are evaluated in Sec.~\ref{sec:uncertainties}, and the final results
are summarized in Sec.~\ref{sec:CONLUSION}.

\section{Detection and Datasets}
\label{sec:dect}
The BESIII detector~\cite{ABLIKIM2010345} records symmetric $\epem$
collisions provided by the BEPCII storage ring~\cite{bepcii}. 
BESIII has collected large data samples in this energy
region~\cite{dataset}.  The cylindrical core of the BESIII detector
covers 93\% of the full solid angle. Starting from the interaction
point (IP), the detector consists of a main drift chamber (MDC), a
time-of-flight (TOF) system, and a CsI(Tl) electromagnetic calorimeter
(EMC), which are all enclosed in a superconducting solenoid magnet
providing a 1.0 T magnetic field.  The solenoid is supported by an
octagonal flux-return yoke with resistive-plate-counter 
muon-identification modules interleaved with steel.  The charged-particle
momentum resolution at 1~GeV/$c$ is 0.5\%, and the $\dedx$ resolution
is 6\% for electrons from Bhabha scattering.  The EMC measures photon
energies with a resolution of 2.5\% (5\%) at 1 GeV in the barrel (end
cap) region.  The time resolution in the TOF barrel region is 68 ps,
and that in the end cap, which was upgraded in 2015 using multigap
resistive plate chamber technology, is 60 ps~\cite{detector2015}.

Simulated data samples are produced with an 
MC framework based on the {\sc geant4} tookit~\cite{sim}, which includes 
a geometric description of the BESIII
detector and detector response.  The simulation uses the {\sc
  kkmc}~\cite{KKMC} generator that takes into account the beam-energy
  spread and initial-state radiation in $\epem$
  annihilations.  Used for background study, the inclusive MC sample
  includes the production of $\epem\to D_s^{*\pm} D_s^{\mp}$, and other open-charm
  processes, as well as the production of vector
  charmonium(like) states and lighter $\qqbar$ pairs 
  incorporated in {\sc kkmc}.  
  All particle decays are modeled with {\sc evtgen}~\cite{EvtGen} using branching fractions 
  either taken from the Particle Data Group~\cite{pdg}, when available, 
  or otherwise estimated with {\sc lundcharm}~\cite{LundCharm}.
  Final-state
  radiation from charged final-state particles is incorporated
  using {\sc photos}~\cite{FSR}.

We also generate an MC sample of $\epem\to D_s^{*\pm} D_s^{\mp}$, 
where one of the two $D_s^{\pm}$ mesons
in each event decays inclusively, and the other 
decays into the signal mode
uniformly distributed in the available
phase space (PHSP), resulting in a uniformly populated
Dalitz plot. This PHSP MC sample is used to evaluate the signal
efficiency as a function of position on the two-dimensional Dalitz plot plane. 

\section{Event Selection}
\label{chap:event_selection}
The selection criteria are based on the reconstruction of one $\Ds\to \pi^+\pi^-\pi^+$
candidate and one photon from the process $\epem\to  D_s^{*\pm} (\to D_s^{\pm}\gamma) D_s^{\mp}$, 
and other $D_s^{*\pm}$ decay modes are ignored. 
The $\Ds$ candidate is
therefore produced either directly from the $\epem$ collision (``direct
``$\Ds$''), or from $\Dss$ decay (``indirect $\Ds$'').  

To reconstruct $\Ds\to\pipipi$, we require that three
charged-track candidates detected in the MDC must 
satisfy $\vert\!\cos\theta\vert<0.93$, where the polar angle
$\theta$ is defined in the laboratory frame with respect to the $z$ axis, which is the
symmetry axis of the MDC. The distance of closest approach to the IP 
is required to be less than 10 cm along the $z$ axis 
and less than 1 cm in the transverse plane 
in the laboratory frame.

Charged tracks are identified as pions or kaons with particle
identification (PID), which combines measurements of $\dedx$
in the MDC and the flight times in the TOF to form likelihoods
${\cal L}(h)$ for different hadron hypotheses $h$ ($h=K,\pi$). Charged
pions are identified by requiring ${\cal  L}(\pi) > {\cal L}(K)$.

Photon candidates in $D_s^{*\pm} \to D_s^{\pm}\gamma$ 
are reconstructed using showers in the EMC.  The
deposited energy of each shower must be more than 25 MeV in the barrel
region ($\vert\!\cos\theta\vert < 0.80$) and more than 50 MeV in the
end cap region ($0.86<\vert\!\cos\theta\vert<0.92$).  To exclude
showers that originate from charged tracks, the angle between the
position of each shower in the EMC and the closest extrapolated
charged track must be greater than $10\degree$ in the laboratory frame.  Further, the
difference between the EMC time and the event start time is required
to be within [0,700]~ns to suppress electronic noise and showers
unrelated to the event.

To identify photons from $D_s^{*\pm} \to D_s^{\pm}\gamma$, 
we require that the
photons are not from any of reconstructed neutral pions. In a
$\piz\to\gamma\gamma$ reconstruction, the photon pair is required to
satisfy the above photon selection criteria.  The $\pi^0$ candidate
is selected with a requirement on the invariant mass of the pair of
$0.125<m(\gamma\gamma)<0.145$~GeV/$c^2$.  The requirement on the photons
loses
$\sim$10\% of signal $\Ds\to\pipipi$ candidates in which $\sim$60\%
have an accompanying photon 
that is in
fact from a $\piz$.

To suppress pion contributions from $\KS\to \pipi$ and simplify the modeling 
of the $\pipi$ invariant mass spectra in the amplitude analysis, we also
reconstruct $\KS$ candidates from two oppositely charged tracks each
with the distance of closest approach to the IP less than 20 cm along the $z$ axis.
The two charged tracks are
assigned as $\pipi$ without imposing any PID criteria.  They are
constrained to originate from a common vertex and required to have an
invariant mass within 12~\mevcc of  
the known $K^0$ mass~\cite{pdg}. The decay length
%
of the $\KS$ candidate 
is required to be greater than twice the vertex
resolution.  Any charged pion candidate that is found
to be also part of a reconstructed $\KS$ is rejected. We
reduce $\Ds\to \KS(\to\pipi) \pip$ backgrounds to a negligible level
($\sim$0.4\% of signals) and retain 98\% of $\Ds\to\pipipi$ signal candidates 
that are free of any $\KS$ contribution.

For each $\Ds\to\pipipi$ candidate, we require the three-pion invariant
mass $m(\Ds)$ to be between 1.9000 and 2.0535~$\gevcc$. We define the recoil mass of $\Ds$ as

\begin{equation}
M_{\rm rec}c^2 = \sqrt{\left(E_{\rm c.m.} - \sqrt{\left|\vec{p}_{\Ds}c\right|^2+m^2_{D_s}c^4}\right)^2-\left|\vec{p}_{\Ds}c\right|^2}\,,
\end{equation}
where $\vec{p}_{\Ds}$ is the reconstructed momentum of the $\Ds$
candidate (sum of the momenta of the three pions) in the $\epem$
center-of-mass frame and $m_{D_s}$ is the known \Ds\ mass~\cite{pdg}.
For direct $\Ds$
candidates, the $M_{\rm rec}$ distribution will peak around the known
$\Dss$ mass $m_{D_s^*}$, and for indirect $\Ds$ candidates, the mass
difference $\Delta M \equiv m(\Ds\gamma)-m(\Ds)$ will peak around the
known mass difference of $\Dss$ and $\Ds$~\cite{pdg}. In the 
presence of multiple photon candidates in an event, we identify the
$\gamma$ from $\Dsspm$ decay by selecting the one that gives the recoil mass
of the $\pipipi\gamma$ system closest to $m_{D_s}$.

To suppress the combinatorial background formed by random combinations of tracks, 
a multivariate classifier based on the Multilayer perceptron implementation
of artificial neural networks
(NNs) from the TMVA package~\cite{tmva} 
is used. Trained on simulated $\Ds\to\pipipi$
 and inclusive MC samples with signal decays removed as signal samples and background
samples respectively, the classifiers use 
different sets of input
parameters for the two categories depending on the $\Ds$ origin as
in Ref.~\cite{BESIIIkkpi}. 
We consider $\Ds$ candidates with $\left|M_{\rm rec} -m_{D_s^*}\right| \leq 0.02~\gevcc$
 as direct $\Ds$, and
use the following NN input parameters:
\begin{enumerate}[label=(\roman*)]
   \item $M_{\rm rec}$; 
    \item $P_{\rm rest}$, defined as the total momentum of 
         the tracks and photon candidates in the rest of event (not part 
        of the $\Ds\to \pipipi$ candidate);
     \item $E_{\gamma}$, defined as the energy of the photon from the $\Dss$;
     \item $M'_{\rm rec}$, defined as the recoil mass of the $\Ds\gamma$ combination, 

         \begin{equation}         
          M'_{\rm rec}c^2 = \sqrt{\left(E_{\rm c.m.} - \sqrt{\left|\vec{p}_{\Ds\gamma}c\right|^2+m^2_{D_s^*}c^4}\right)^2-\left|\vec{p}_{\Ds\gamma}c\right|^2},
         \end{equation}         
        with $\vec{p}_{\Ds\gamma}$ as the momentum of $\Ds\gamma$; 
        \item and $N_{\rm total}$, defined as the total number of charged tracks and photon candidates in the event.
\end{enumerate}
For indirect $\Ds$, candidates with $\left|M_{\rm rec}
-m_{D_s^*}\right| > 0.02~\gevcc$, and $0.135 <\Delta M <
0.15$~\gevcc are considered. The following NN input parameters are used in this case:
\begin{enumerate}[label=(\roman*)]
        \item $\Delta M$; 
        \item $P'_{\rm rest}$, defined as the total momentum of 
            the tracks and photon candidates in the rest of event (not part 
        of the $\Dss\to\Ds\gamma,\ \Ds\to \pipipi$ combination); 
        \item $M'_{\rm rec}$;
        \item and $N_{\rm total}$.
\end{enumerate}

\begin{figure*}[!htb]
\centering
    \subfigure{
\includegraphics[width=0.4\linewidth]{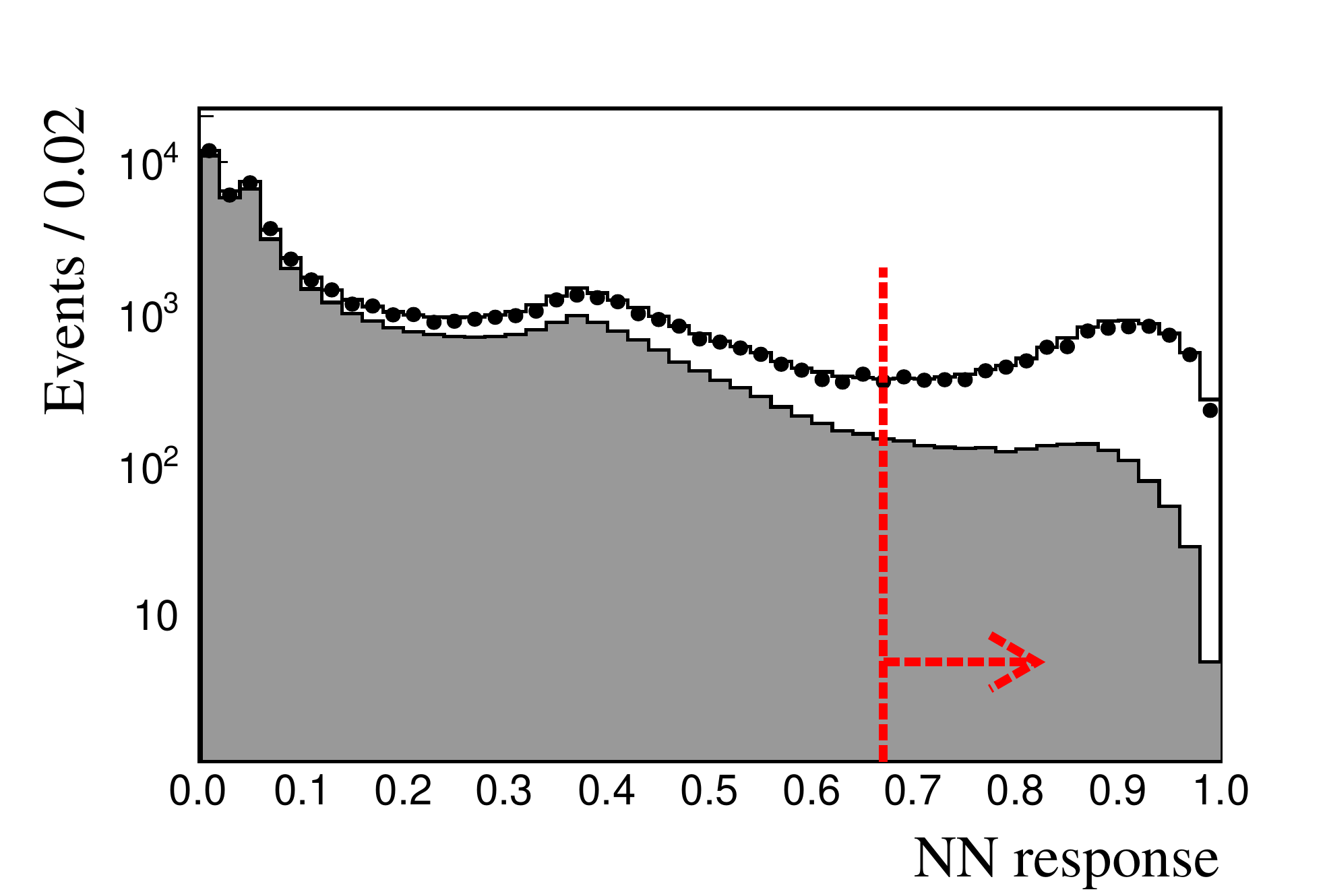}
    \putat{-150}{+100}{(a)}
    }
    \subfigure{
\includegraphics[width=0.4\linewidth]{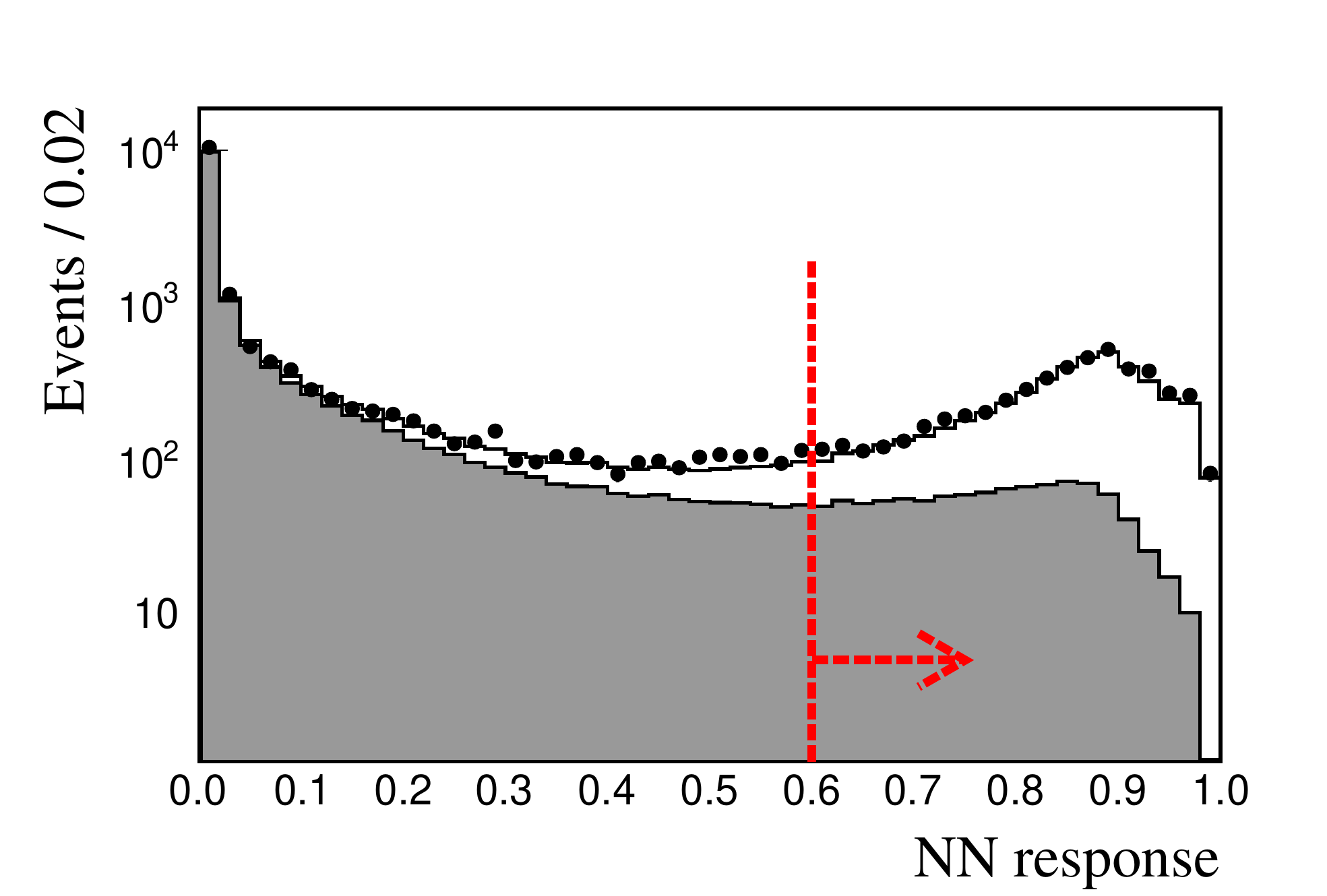}
    \putat{-150}{+100}{(b)}
    }
    \caption{NN response distributions for (a) direct $\Ds$ and (b)
      indirect $\Ds$ categories for data and inclusive MC
      simulation.  The data histograms (points) are compared to MC simulated
      histograms (lines) including both signal and background (shaded areas)
      contributions. MC histograms are scaled based on the integrated luminosity of data. 
      The
      requirements on the NN responses are marked by the vertical dashed
      lines in red. The $\chi^2$ test~\cite{chi2test} results for comparison of the data and weighted MC histograms are also shown with ${\rm N}_{\rm bins}$ as the number of histogram bins.}
\label{fig:mlpdatamc}
\end{figure*}

Our NN response distributions in data and MC samples are compared in
Fig.~\ref{fig:mlpdatamc} for both $\Ds$ categories, where good
agreement is observed. 
With the NN response requirements also shown in Fig.~\ref{fig:mlpdatamc}, 
 we are able to achieve a signal purity of about 80\% in the signal region 
($\vert\!m(\Ds)-m_{D_s}\vert<12$~\mevcc).
The NN response requirements are chosen in order to be in line with the BABAR analysis with similar amount of data~\cite{babarpaper}. 

\begin{figure}[!htb]
\centering
\includegraphics[width=0.9\linewidth]{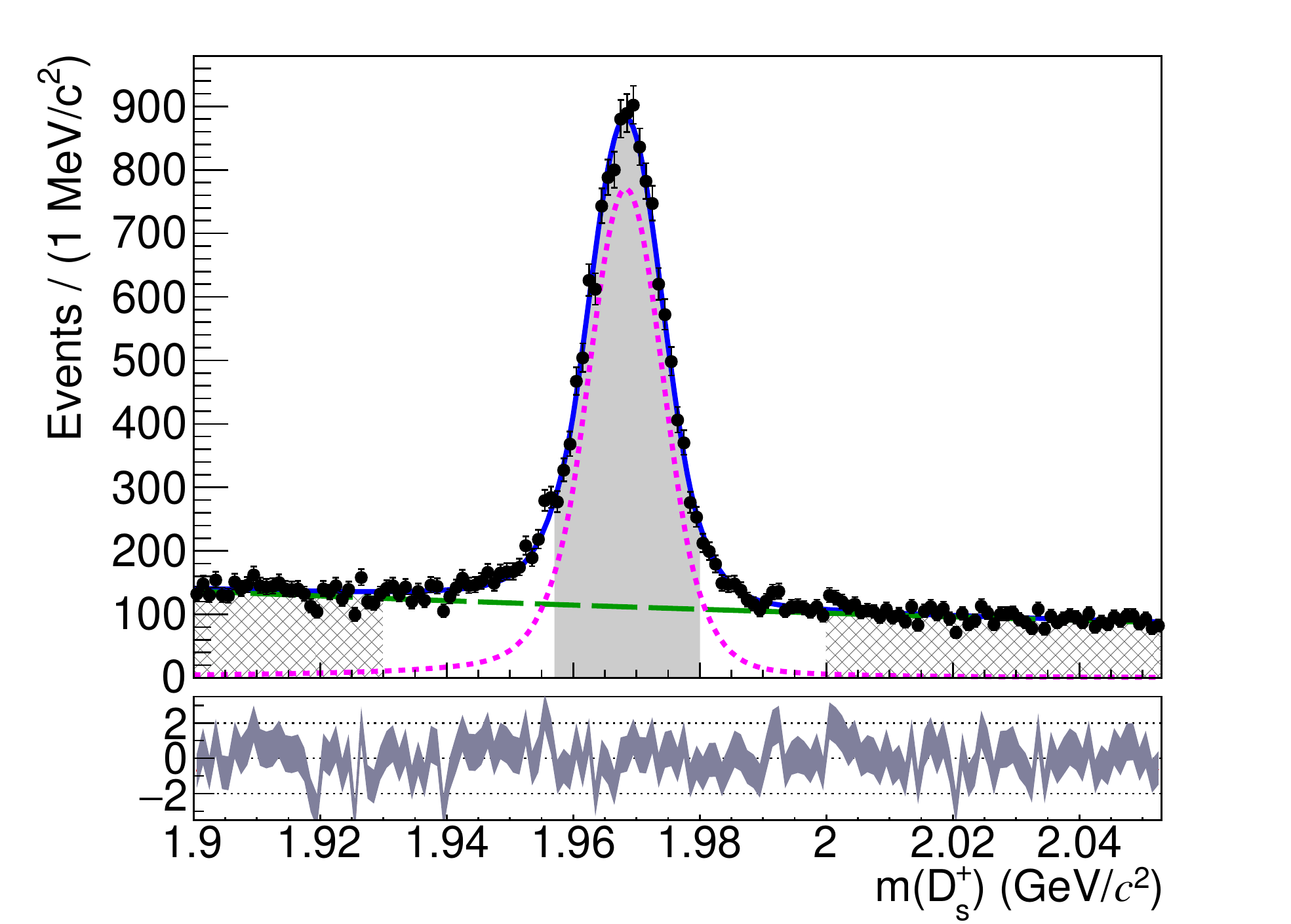}
    \caption{The invariant mass distribution of $\Ds\to\pipipi$
      candidates of data. 
      Data (points) are shown together with the total fit
      (blue), signal PDF (magenta dashed), and background PDF (light
      green long dashed).  The 
      signal region corresponds
      to the shaded region, and the sideband events are taken from
      the cross-hatched regions.  }
\label{fig:mDsfit}
\end{figure}

With
the NN response requirements applied, we perform an unbinned maximum
likelihood fit to the data distribution of $m(\Ds)$, as shown in Fig.~\ref{fig:mDsfit} to determine the signal purity within 
the signal region to be $(80.6\pm 1.0)$\% where the uncertainty is statistical only.  The
background is modeled by an exponential probability density function (PDF), and the signal is
modeled by the sum of a Gaussian PDF and a double-sided crystal-ball
(DSCB) PDF~\cite{dscb} with a common mode value. The DSCB PDF tail
parameters, as well as the ratio of the width parameters from the DSCB and
Gaussian PDFs, are determined from the PHSP MC sample and fixed in the fit to
data. 

\begin{figure}[!htb]
\centering
\includegraphics[width=0.8\linewidth]{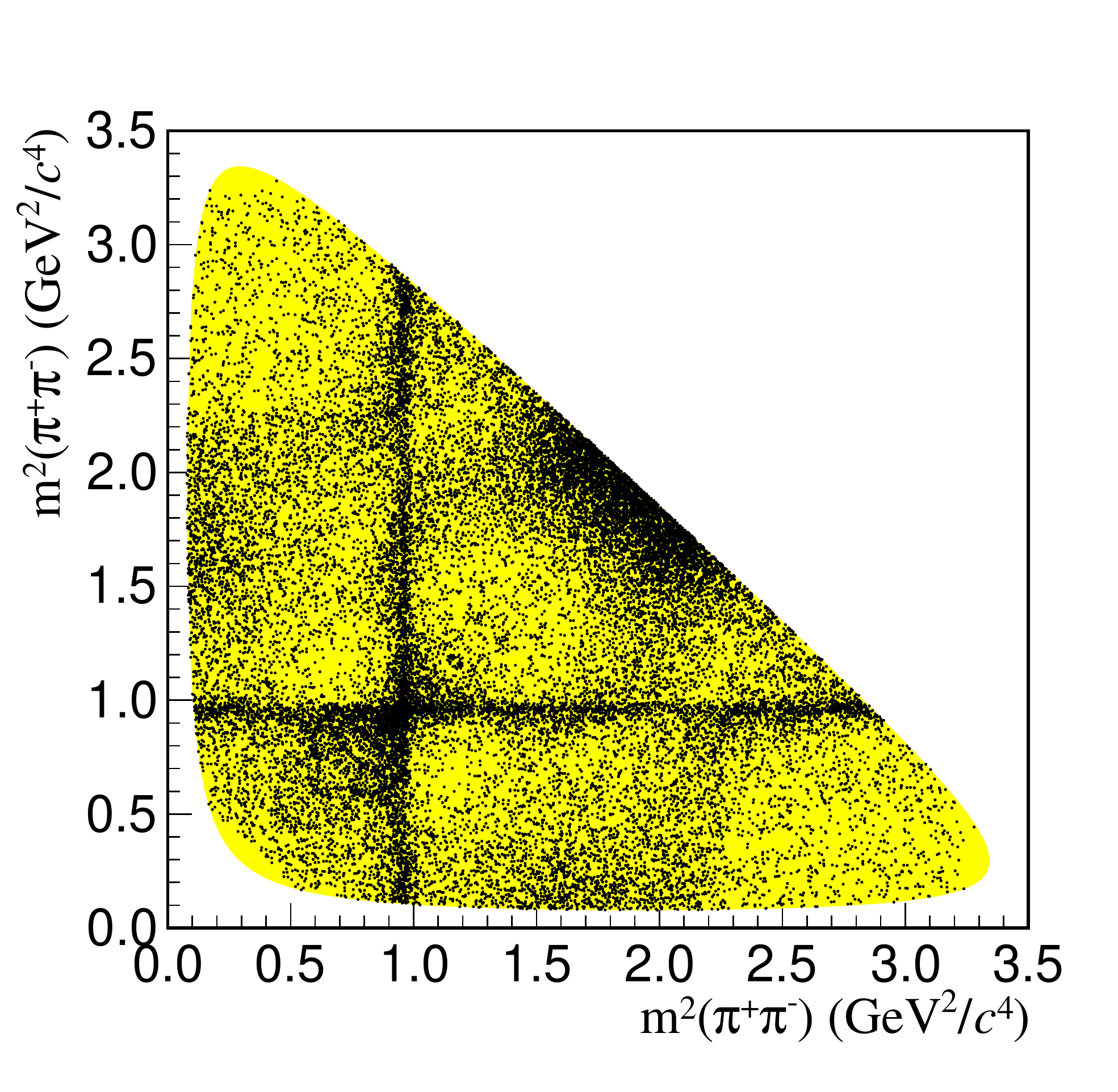}
    \caption{Symmetric Dalitz plot of $\Ds\to \pipipi$ from
      data within the signal region. 
      All individual events are shown, 
      with two entries per $\Ds$ candidate. The yellow area represents the phase
      space available to the decay.
    }
\label{fig:dpdata}
\end{figure}

Finally, we perform a kinematic fit to all $\Ds\to\pipipi$ candidates which enforces a $D_s^+$ mass~\cite{pdg}
constraint. The kinematic-fit-corrected four-momenta of all three pions of $\Ds$ are 
used to calculate $\pi\pi$ invariant masses for the following amplitude analysis.
In total we have 13,797 data events within the signal
$\Ds$ mass region. This is slightly
more (5\%) than BABAR\ with roughly the same 80\%
signal purity. Using the track momenta obtained from the $\Ds$ mass-constrained kinematic fit,
the Dalitz plot symmetrized over particle content 
for this sample is shown in Fig.~\ref{fig:dpdata},
where both possible entries from one $\Ds$ candidate, ($m^2(\pip\pim)_{\rm low}$, $m^2(\pip\pim)_{\rm high}$) and ($m^2(\pip\pim)_{\rm high}$, $m^2(\pip\pim)_{\rm low}$), are plotted. Here the ``low'' (``high'') subscript marks the lower (higher) value of the two $\pip\pim$ mass combinations.
A pair of narrow crossing bands corresponding to the $\Ds\to f_0(980)\pip$ process can be clearly seen. Furthermore, concentration of events around the $m^2(\pip\pim) = 1.9$~\gevcc region is visible, 
which hints the presence of broad resonances such as $f_2(1270)$ and $f_0(1370)$.

\section{Amplitude Analysis}
\label{sec:ampana}
This analysis will determine the intermediate-state composition of
$\Ds\to\pipipi$ by analyzing the Dalitz plot of
$\Ds\to\pipipi$ as illustrated in Fig.~\ref{fig:dpdata}.  We randomly assign
$m^2(\pipi)$ of the two $\pipi$ combinations in $\Ds\to\pipipi$ as $x$
and $y$ axes of the Dalitz plot respectively, and the $z$ ($\equiv
m^2(\pip\pip)$) axis is used later in efficiency modeling.

\subsection{Analysis formalism}
\label{anaform}
The decay amplitude of $\Ds\to\pipipi$ is modeled by a coherent sum of
various 
amplitudes 
with
angular momentum quantum numbers $L=0$, $1$, and $2$. Each amplitude
$A_i$ denoted by $i$
is symmetrized with respect to the two identical pions in the decay,

\begin{equation}
    {\cal A}_i(x,y) = A_i(x, y) + A_i(y, x). \label{eq:Amp}
\end{equation}

In our QMIPWA formalism, which is similar to that used by the Fermilab
E791 Collaboration to model the $K\pi$ \swave~\cite{mipwae791}, the
$\pipi$ \swavehy\ amplitude is a complex function of 
$m(\pipi)$.  Using the same choice of dividing the $m(\pipi)$ spectrum as in
Ref.~\cite{babarpaper}, the \swavehy\ amplitude $A_0(x,y)$ at 29 control points
each with an index $k$ 
is modeled by two real parameters $a_k$ (magnitude) and
$\gamma_k$ (phase). A cubic spline interpolation is used to
 parametrize 
both the real and imaginary parts of $A_0(x,y)$ 
for $x_k \leq x < x_{k+1}$.

Using the isobar model for ${\cal P}$ waves and ${\cal D}$ waves, the full amplitude 
is written as
a coherent sum of amplitudes ${\cal A}_i$ with complex coefficients $c_i$, 

\begin{equation}
    {\cal A}(x,y) = \sum_i c_i {\cal A}_i (x,y)\,,
    \label{eq:dpamp_pwa}
\end{equation}
where $c_0\equiv 1$, so the free parameters to model the \swavehy\ amplitude remain unchanged, and the other coefficients
$c_i\equiv |c_i| e^{i \phi_i}$ for ${\cal P}$- or ${\cal D}$-wave amplitudes.
Each ${\cal P}$- or ${\cal D}$-wave amplitude 
is represented 
by the product of Blatt-Weisskopf barrier factors $F_{D_s,r}^L$~\cite{barrier}, a complex relativistic Breit-Wigner function $W^L$ and a real spin-dependent angular term $Z^L$,

\begin{equation}
    A_i(x,y) = F_D^L(x,y) F_r^L(x,y) W^L(x) Z^L(x,y)\,.
    \label{eq:dpamp_2}
\end{equation}
The detailed formalism is the same as the one in Ref.~\cite{babarDsKKpi} and therefore left out in the paper.
For the Blatt-Weisskopf barrier factors, we set the radii of $\Ds$ and intermediate
resonances to be $R_{D_s}=5$~${(\gevc)}^{-1}$ and $R_r= 1.5$~${(\gevc)}^{-1}$, respectively.

\subsection{Efficiency}
\label{sec:eff}
To  
model the efficiency across the Dalitz plot plane for signal events,
we perform an unbinned maximum likelihood fit to the PHSP MC sample with
a function
$\eta(x,y) = {\cal P}(x,y){\cal T}(x){\cal T}(y){\cal T}(z)$
that is symmetric under reflection through $y=x$.
Here ${\cal P}(x,y)$ is a
two-dimensional polynomial function centered on an arbitrary point 
($x_c$, $y_c$) = (1, 1)~GeV$^2$/$c^4$ on the Dalitz plot plane, 

\begin{multline}
\label{eqn:efficiency}
    {\cal P}(x,y) = 1 + E_1\left(\hat{x}+\hat{y}\right)
                      + E_2\left(\hat{x}^2+\hat{y}^2\right)
                      +E_{11}\hat{x}\hat{y}\\
                      + E_3\left(\hat{x}^3+\hat{y}^3\right)
                      + E_{12}\left(\hat{x}^2\hat{y} + \hat{x}\hat{y}^2\right),
\end{multline}
where $\hat{x} = x-x_c$, 
$\hat{y} = y-y_c$. 
${\cal T}(v)$ is a sinelike threshold function for each
Dalitz plot variable, $v$  ($\equiv x, y$ or $z$),

\begin{equation}
\label{eqn:threshold_factor}
    {\cal T}(v) = \left\{
\begin{array}{ll}                   \sin( E_{{\rm th},v}\cdot|v-v_{\rm max}| ),
    & E_{{\rm th},v}\cdot|v-v_{\rm max}|<\tfrac{\pi}{2}, \\
                 1              , & E_{{\rm th},v}\cdot|v-v_{\rm max}| \geq \tfrac{\pi}{2}, \\
\end{array} \right.
\end{equation}
where all polynomial coefficients,
$E_1$, $E_{2}$, $E_{11}$, $E_{3}$, $E_{12}$, 
and $E_{{\rm th},v}$
are the fit parameters (requiring $E_{{\rm th},y} \equiv E_{{\rm th},x}$).
Each variable $v$ has one threshold, $v_{\rm max}\equiv \left(m_{D_s}-m_{\pi}\right)^2$, the kinematic limit of $m^2(\pi\pi)$, where $m_{\pi}$ is the known mass
 of $\pip$. 
The fit parameters for the efficiency function $\eta(x,y)$ determined in the fit are 
shown in 
Table~\ref{tab:effpar}. 
These parameters are fixed in fits to data, and their associated uncertainties are later considered as a source of systematic uncertainties. 


\begin{table}
\caption{Fit results of the efficiency function $\eta(x,y)$ from the PHSP MC sample.}
\begin{center}
    \begin{tabular}{
    l|c}
        \hline\hline Parameter & Value  \\\hline
        $E_1$ & $\phantom{-}0.064 \pm 0.003$ \\
    $E_2$ & $-0.066\pm 0.004$\\
    $E_3$ & $-0.006\pm 0.002$ \\
    $E_{11}$ & $-0.158\pm 0.006$ \\
    $E_{12}$& $\phantom{-}0.090\pm 0.006$ \\ \hline
    $E_{{\rm th},x(y)}$ &  $\phantom{-}1.516 \pm 0.019$ \\
    $E_{{\rm th},z}$ & $\phantom{-}1.563 \pm 0.028$ \\ \hline\hline
\end{tabular}
\label{tab:effpar}
\end{center}
\end{table}


\subsection{Likelihood function construction}
\label{sec:nll}

We perform
an unbinned maximum-likelihood fit to the distribution of $\Ds$ candidates in the Dalitz plot. The likelihood function is 

\begin{multline}
    {\cal L} = \prod_{{\rm events}}\left\{F\left(m_j(\Ds)\right)\eta\left(x_j,y_j\right)\frac{\left|{\cal A}\left(x_j,y_j\right)\right|^2}{\int_{\rm DP} \left|{\cal A}\left(x,y\right)\right|^2\eta\left(x,y\right)dxdy}\right.\\ 
    \left.+\left[1-F\left(m_j(\Ds)\right)\right]{\cal B}(x_j,y_j)\vphantom{\frac{\sum_{i,j}}{\sum_{i,j}}}\right\}\,,\label{eq:LH}
\end{multline}
where 
\begin{enumerate}[label=(\roman*)]
    \item $j$ is a $\Ds$ candidate index.    
    \item $F\left(m(\Ds)\right)$ is the signal fraction, depending on $m(\Ds)$ 
        before the kinematic fit mentioned in Sec.~\ref{chap:event_selection}. It is defined as

        \begin{equation*}
        F\left(m(\Ds)\right) = \frac{{ S}\left(m(\Ds)\right)}{{ S}\left(m(\Ds)\right)+{ B}\left(m(\Ds)\right)}\,,
        \end{equation*}
        where ${ S}$ and ${ B}$ are the signal and background functions determined from
        fitting to the data mass distribution as depicted in Fig.~\ref{fig:mDsfit}.
    \item ${\cal B}(x,y)$ is the background function, which is modeled by a histogram taken from $\Ds$ candidates
        in the data sideband [$1.90<m(\Ds) < 1.93~\gevcc$ and $2.00<m(\Ds) < 2.0535~\gevcc$, as shown in Fig.~\ref{fig:mDsfit}]. We also have the normalization requirement $\int {\cal B}(x,y)dxdy = 1$.
    \item DP is the integral limit denotes the the kinematic limit of the Dalitz plot. The integral is
        calculated numerically based on a large number of PHSP MC
        events at the generator level, that is, without the simulation on
detector responses.
\end{enumerate}
The fit fraction for the $i$th signal amplitude 
is defined as 

\begin{equation}
    {\cal F}_i = \frac{|c_i|^2\int |{\cal A}_i\left(x,y\right)|^2 dx dy}{\int \left|{\cal A}\left(x,y\right)\right|^2dxdy}\,. 
    \label{eq:ff}
\end{equation}
Statistical uncertainties on the fractions include uncertainties on both magnitudes and phases of different signal amplitudes, 
and are computed using the full covariance matrix. 

\subsection {Fitting}
\label{gpu}

Due to the large number of $\Ds$ candidates and many parameters involved for the
\swavehy\ parametrization during fitting the Dalitz plot model to the data, an open-source
framework called {\tt GooFit}~\cite{goofit} has been used to speed up
the fitting using the parallel processing power of graphical
processing units.  The cubic spline interpolation method is
implemented based on the {\tt GooFit} framework~\cite{Sun:2017wtf}.

\section{Results of the MIPWA}
\label{fitting}
\begin{figure*}[!htb]
\centering
    \subfigure{
    \includegraphics[width=0.70\linewidth]{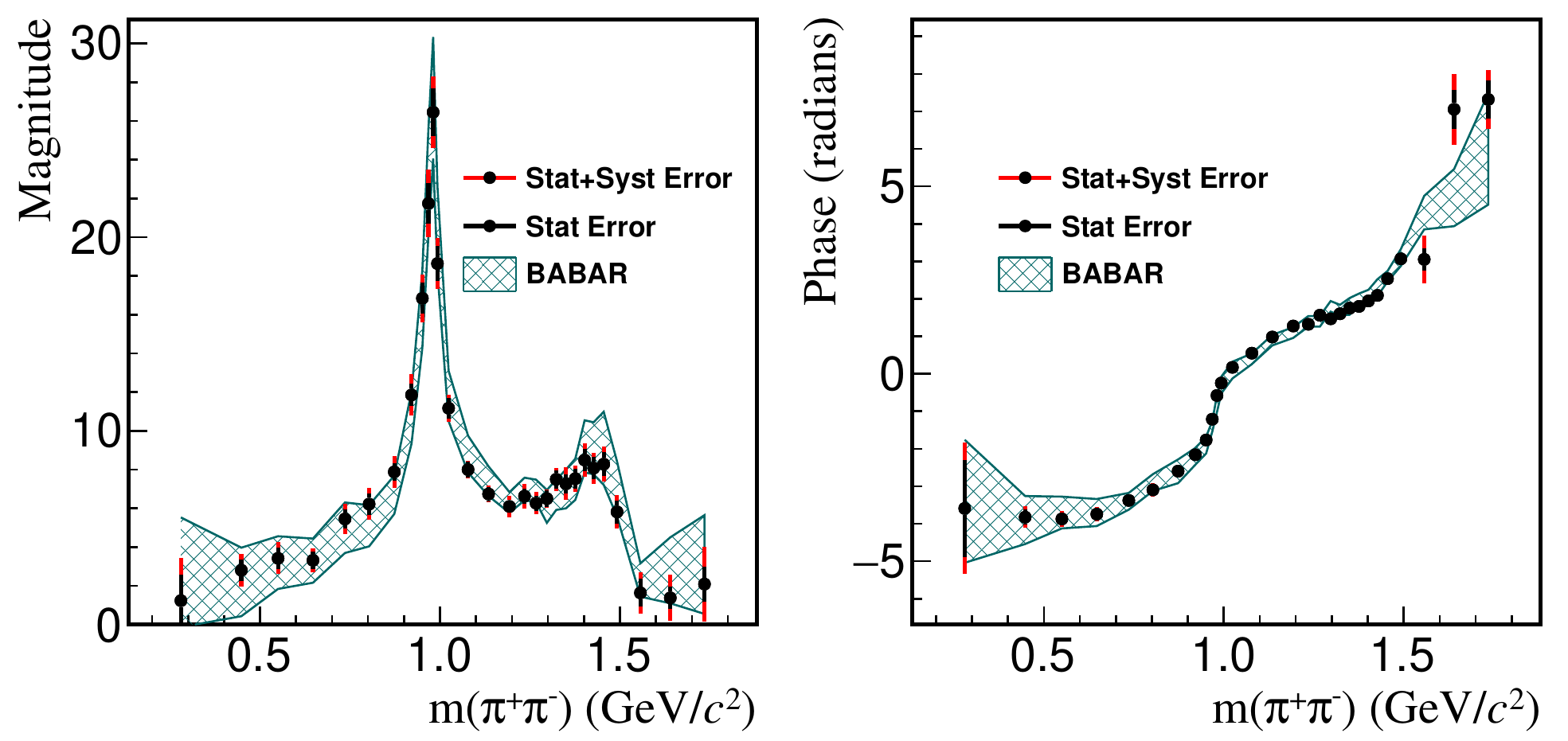}
    \putat{-320}{+140}{(a)}
    \putat{-140}{+140}{(b)}
    }
    \caption{(a) Magnitudes and (b) phases of the \swavehy\ control points as summarized in Table~\ref{tab:fitres_sw}. The results are compared to the BABAR results~\cite{babarpaper} with the same choice on the control points and similar data sample size.}
\label{fig:swaveresult}
\end{figure*}

As in the previous BABAR analysis with similar data sample size, our baseline
signal model includes three intermediate resonances with $L \neq
0$: two ${\cal P}$ waves ($\rho(770)$ and $\rho(1450)$) and one ${\cal
  D}$ wave ($f_2(1270)$). The partial wave with $f_2(1270)$ is the
reference amplitude with the magnitude and phase fixed at 1 and 0, respectively.  
With the masses and widths of these three
resonances fixed to the world averages, and accounting for the
magnitude and phase for each of the two $\rho$ resonances and
the 29 \swavehy\ control points relative
to the reference amplitude, our
baseline signal model contains $N_{\rm par} = 62$ free fit parameters.
Tables~\ref{tab:fitres} and~\ref{tab:fitres_sw} summarize the results
from the amplitude analysis, while the statistical and systematic correlation matrices
 are given in the Appendix and the Supplemental Material~\cite{supp}.
The fit fraction results in Table~\ref{tab:fitres} 
show a clear domination of $\pipi$ \swavehy contribution. Also, 
a notable ${\cal D}$-wave contribution from $\Ds\to f_2(1270)\pip$ is found,
 and the ${\cal P}$-wave contributions from $\rho$ are considerably smaller. 
Our \swavehy\ results are also shown in Fig.~\ref{fig:swaveresult},
where at least one resonance around 1~\gevcc is clearly visible, hinting the presence
of $f_0(980)$, and contributions from other higher mass scalars such as $f_0(1370)$.
The comparison with the BABAR measurements shown in Fig.~\ref{fig:swaveresult}
indicates generally good agreement.

\begin{figure*}
\centering
    \subfigure{
    \includegraphics[width=0.40\linewidth]{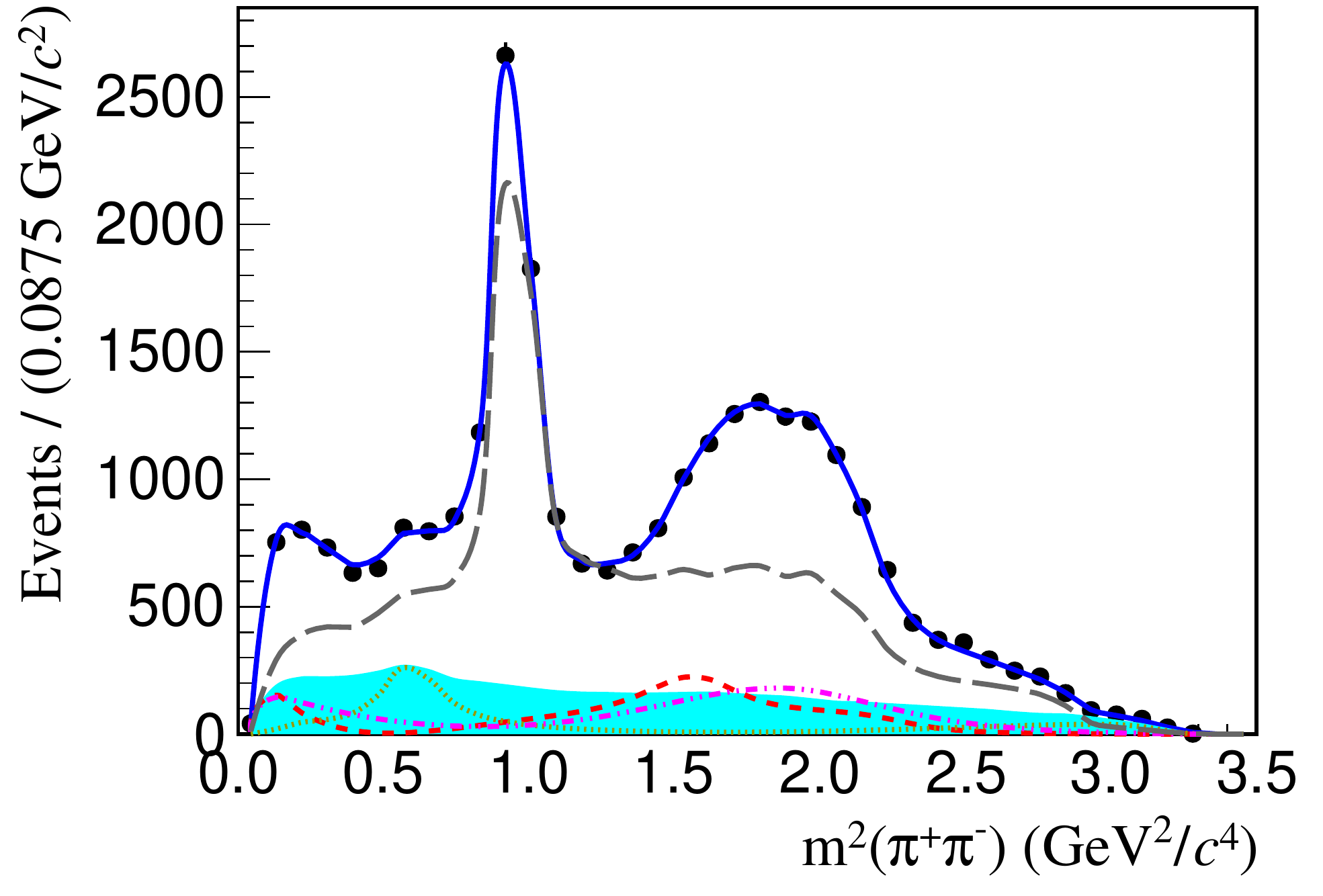}
    \putat{-150}{+120}{(a)}}
    \subfigure{
    \includegraphics[width=0.40\linewidth]{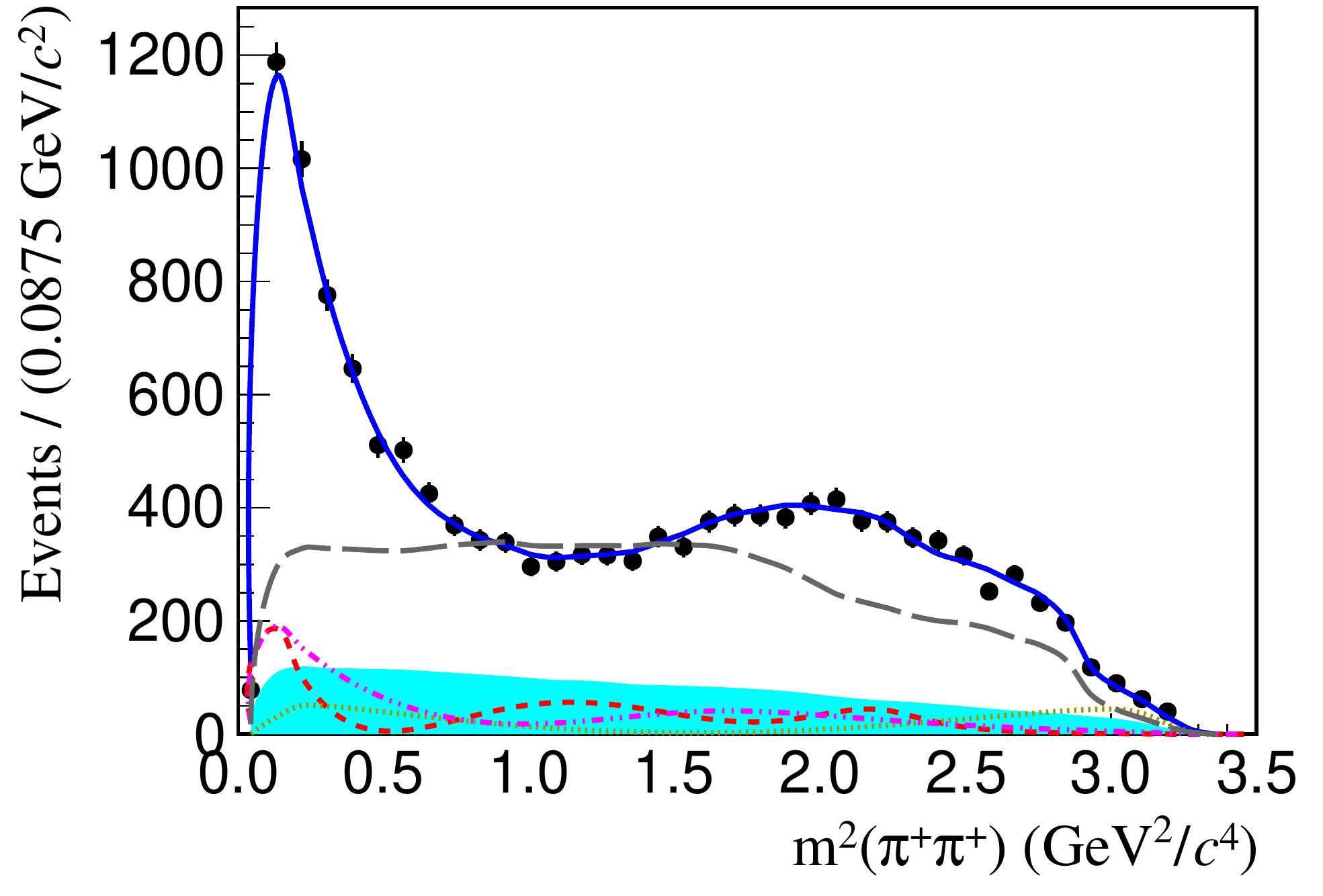}
    \putat{-150}{+120}{(b)}}\\
    \subfigure{
    \includegraphics[width=0.40\linewidth]{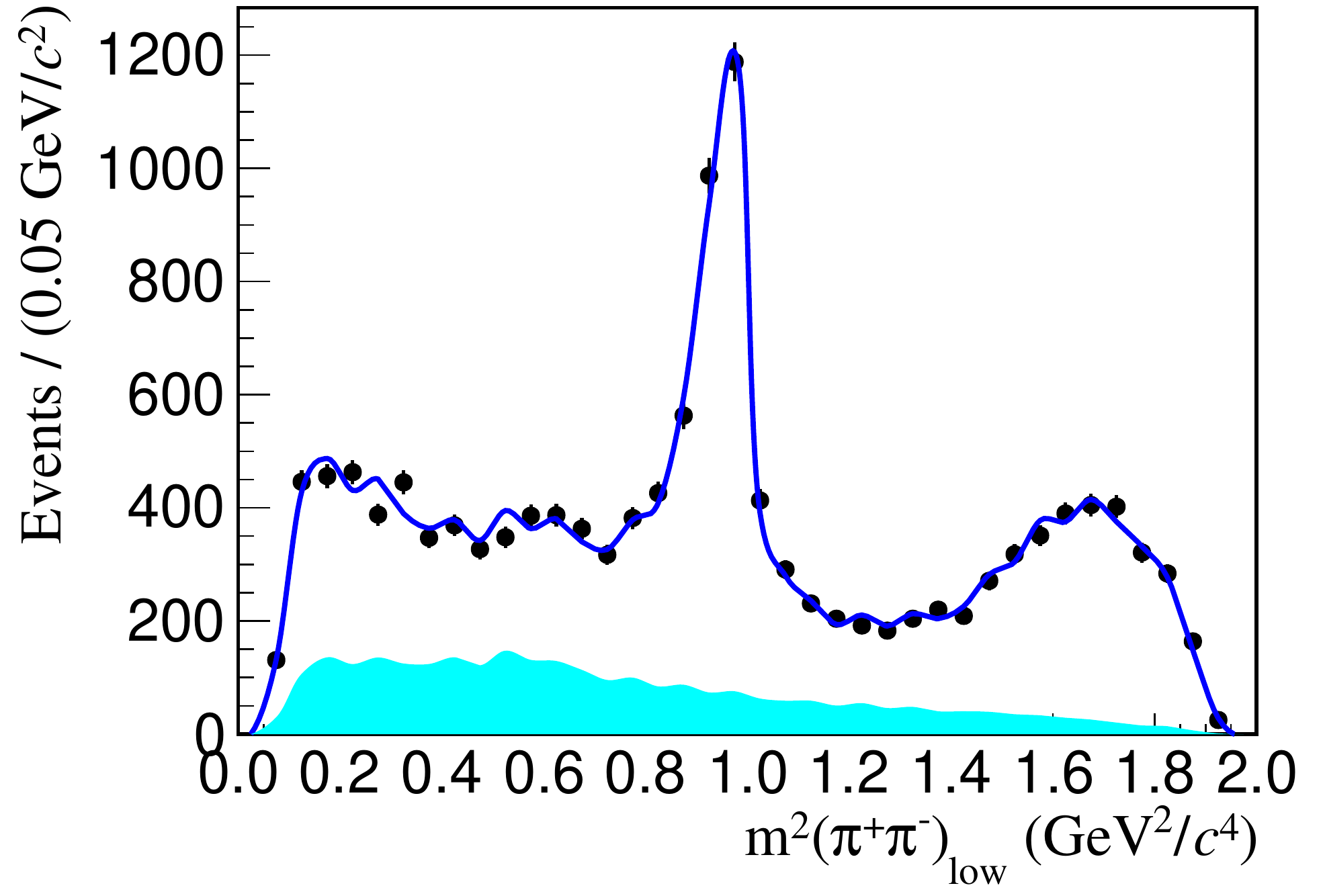}
    \putat{-150}{+120}{(c)}}
    \subfigure{
    \includegraphics[width=0.40\linewidth]{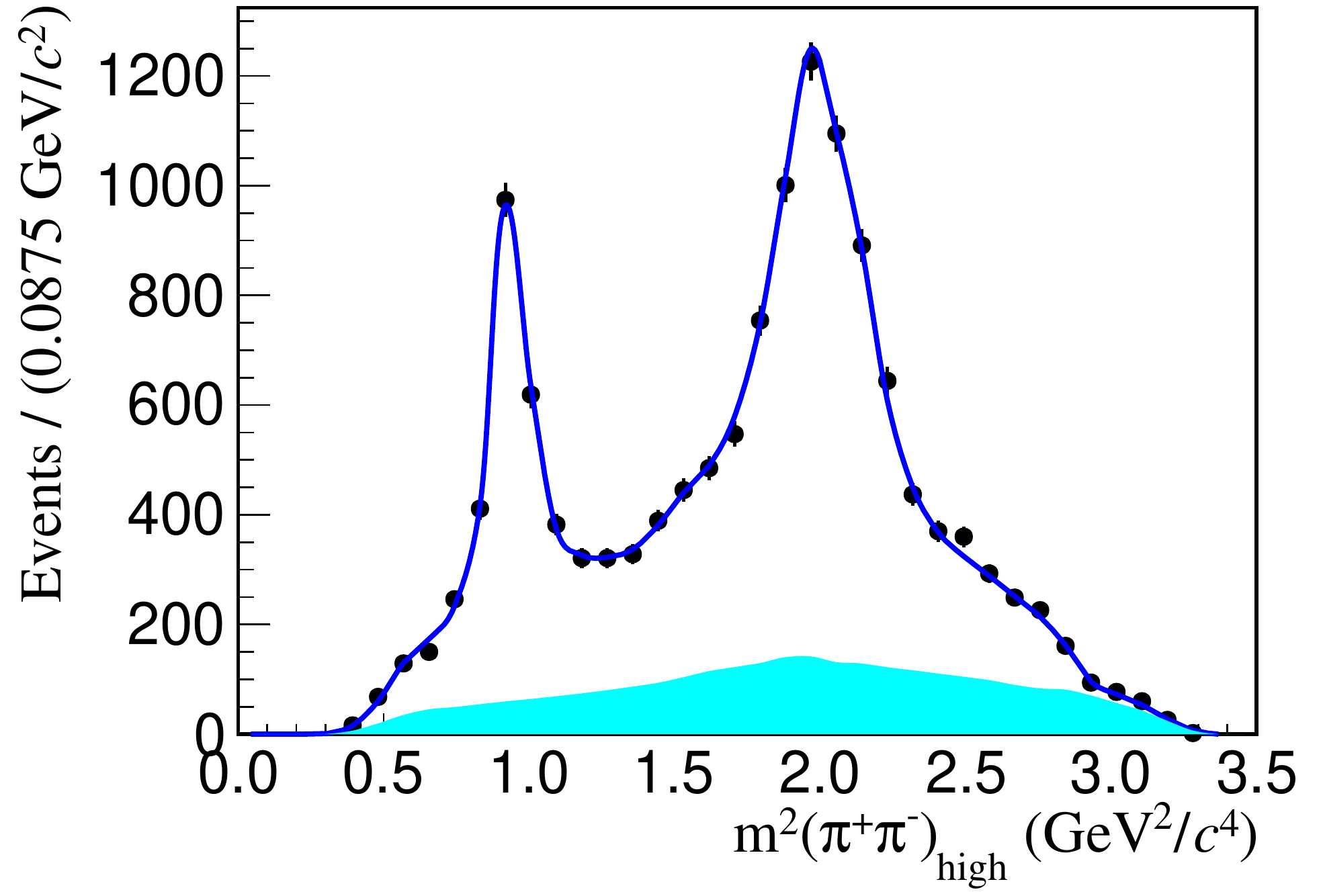}
    \putat{-150}{+120}{(d)}}\\
    \caption{ Projections of data (points with error bars)
      and total fit results (blue lines) on (a) total
      $m^2(\pi^+\pi^-)$ (including both $\pip\pim$ combinations in a $\Ds$ candidate), (b) $m^2(\pi^+\pi^+)$, (c) low-mass
      combination $m^2(\pi^+\pi^-)_{\rm low}$, and (d) high-mass
      combination $m^2(\pi^+\pi^-)_{\rm high}$. The shaded areas in
      cyan are the background contributions. Also shown in (a)
      and (b) are contributions from the \swave\ (gray long-dashed lines),
      $\rho(770)$ (yellow dotted lines, scaled by a factor of 10 for
      better visibility), $\rho(1450)$ (magenta dot-dashed line,
      scaled by a factor of 10 for better visibility), and $f_2(1270)$
      (red short-dashed lines).  }
\label{fig:nominalfit}
\end{figure*}

Our fit projections are determined by producing a large number of PHSP
MC events at the generator level, 
which are weighted by the
fit likelihood function [Eq.~(\ref{eq:LH})], and normalized (with the
weighted sum) to the observed number of data candidates. 
The fit projections are shown in
Fig.~\ref{fig:nominalfit}, with the data overlaid for comparison. 

\begin{table*}
\centering
    \caption{Fit fractions,
    magnitudes, and phases from our baseline fit. 
    The uncertainties are statistical and systematic, respectively. The bottom line shows the sum of all fit fractions. The sum of the fit fractions is not necessarily equal to unity due to the interferences among the contributing amplitudes.}
\label{tab:fitres}
\begin{tabular}{cccc}
\hline\hline
    Decay mode & Fit fraction (\%) & Magnitude & Phase (radians)\\
\hline    
$f_2(1270)\pip$ & 10.5 $\pm$ 0.8 $\pm$ 1.1& 1. (Fixed) & 0. (Fixed) \\
$\rho(770)\pip$ & 0.9 $\pm$ 0.4 $\pm$ 0.5& 0.13 $\pm$ 0.03 $\pm$ 0.04& 5.44 $\pm$ 0.25 $\pm$ 0.60 \\
$\rho(1450)\pip$ & 1.3 $\pm$ 0.4 $\pm$ 0.5& 0.91 $\pm$ 0.16 $\pm$ 0.21& 1.03 $\pm$ 0.32 $\pm$ 0.32 \\
\swave & 84.2 $\pm$ 0.8 $\pm$ 1.2& Table~\ref{tab:fitres_sw} & Table~\ref{tab:fitres_sw} \\
$\sum_i {\cal F}_i$  & 96.8 $\pm$ 2.4 $\pm$ 3.3 & & \\
    
\hline
\hline
\end{tabular}
\end{table*}

\begin{table*}
\centering
    \caption{Magnitudes and phases of the $\pipi$ \swavehy\ control points from our baseline fit. 
    The uncertainties are statistical and systematic, respectively. }
\label{tab:fitres_sw}
\begin{tabular}{cccc}
\hline\hline
    Point & Mass (\gevcc) & Magnitude & Phase (radians) \\
\hline
1 & 0.280 & $\phantom{0}1.23 \pm 1.34 \pm 1.73$ & $-3.59 \pm 1.29 \pm 1.19$\\
2 & 0.448 & $\phantom{0}2.80 \pm 0.55 \pm 0.62$ & $-3.82 \pm 0.20 \pm 0.21$\\
3 & 0.550 & $\phantom{0}3.42 \pm 0.54 \pm 0.62$ & $-3.87 \pm 0.15 \pm 0.14$\\
4 & 0.647 & $\phantom{0}3.32 \pm 0.46 \pm 0.42$ & $-3.74 \pm 0.15 \pm 0.13$\\
5 & 0.736 & $\phantom{0}5.45 \pm 0.49 \pm 0.60$ & $-3.38 \pm 0.12 \pm 0.09$\\
6 & 0.803 & $\phantom{0}6.22 \pm 0.55 \pm 0.61$ & $-3.10 \pm 0.13 \pm 0.14$\\
7 & 0.873 & $\phantom{0}7.88 \pm 0.46 \pm 0.66$ & $-2.60 \pm 0.12 \pm 0.09$\\
8 & 0.921 & $11.85 \pm 0.57 \pm 0.90$ & $-2.16 \pm 0.12 \pm 0.10$\\
9 & 0.951 & $16.84 \pm 0.80 \pm 0.93$ & $-1.77 \pm 0.11 \pm 0.08$\\
10 & 0.968 & $21.74 \pm 1.05 \pm 1.41$ & $-1.21 \pm 0.11 \pm 0.09$\\
11 & 0.981 & $26.45 \pm 1.23 \pm 1.40$ & $-0.58 \pm 0.11 \pm 0.07$\\
12 & 0.993 & $18.64 \pm 0.89 \pm 0.98$ & $-0.25 \pm 0.10 \pm 0.08$\\
13 & 1.024 & $11.17 \pm 0.55 \pm 0.44$ & $\phantom{-}0.17 \pm 0.10 \pm 0.07$\\
14 & 1.078 & $\phantom{0}8.00 \pm 0.42 \pm 0.17$ & $\phantom{-}0.55 \pm 0.10 \pm 0.05$\\
15 & 1.135 & $\phantom{0}6.74 \pm 0.36 \pm 0.22$ & $\phantom{-}0.98 \pm 0.09 \pm 0.07$\\
16 & 1.193 & $\phantom{0}6.10 \pm 0.32 \pm 0.46$ & $\phantom{-}1.28 \pm 0.09 \pm 0.03$\\
17 & 1.235 & $\phantom{0}6.63 \pm 0.38 \pm 0.53$ & $\phantom{-}1.32 \pm 0.10 \pm 0.03$\\
18 & 1.267 & $\phantom{0}6.27 \pm 0.39 \pm 0.42$ & $\phantom{-}1.56 \pm 0.11 \pm 0.09$\\
19 & 1.297 & $\phantom{0}6.50 \pm 0.42 \pm 0.25$ & $\phantom{-}1.47 \pm 0.10 \pm 0.06$\\
20 & 1.323 & $\phantom{0}7.50 \pm 0.47 \pm 0.38$ & $\phantom{-}1.60 \pm 0.10 \pm 0.06$\\
21 & 1.350 & $\phantom{0}7.27 \pm 0.49 \pm 0.69$ & $\phantom{-}1.75 \pm 0.10 \pm 0.11$\\
22 & 1.376 & $\phantom{0}7.53 \pm 0.51 \pm 0.45$ & $\phantom{-}1.80 \pm 0.10 \pm 0.12$\\
23 & 1.402 & $\phantom{0}8.49 \pm 0.56 \pm 0.68$ & $\phantom{-}1.94 \pm 0.10 \pm 0.07$\\
24 & 1.427 & $\phantom{0}8.08 \pm 0.57 \pm 0.56$ & $\phantom{-}2.09 \pm 0.11 \pm 0.09$\\
25 & 1.455 & $\phantom{0}8.28 \pm 0.63 \pm 0.63$ & $\phantom{-}2.54 \pm 0.09 \pm 0.09$\\
26 & 1.492 & $\phantom{0}5.82 \pm 0.60 \pm 0.61$ & $\phantom{-}3.07 \pm 0.10 \pm 0.12$\\
27 & 1.557 & $\phantom{0}1.64 \pm 0.72 \pm 0.79$ & $\phantom{-}3.05 \pm 0.30 \pm 0.56$\\
28 & 1.640 & $\phantom{0}1.38 \pm 0.57 \pm 1.06$ & $\phantom{-}7.06 \pm 0.52 \pm 0.78$\\
29 & 1.735 & $\phantom{0}2.09 \pm 0.89 \pm 1.70$ & $\phantom{-}7.32 \pm 0.51 \pm 0.60$\\
\hline
\hline
\end{tabular}
\end{table*}

\begin{figure}
\centering
    \includegraphics[width=0.8\linewidth]{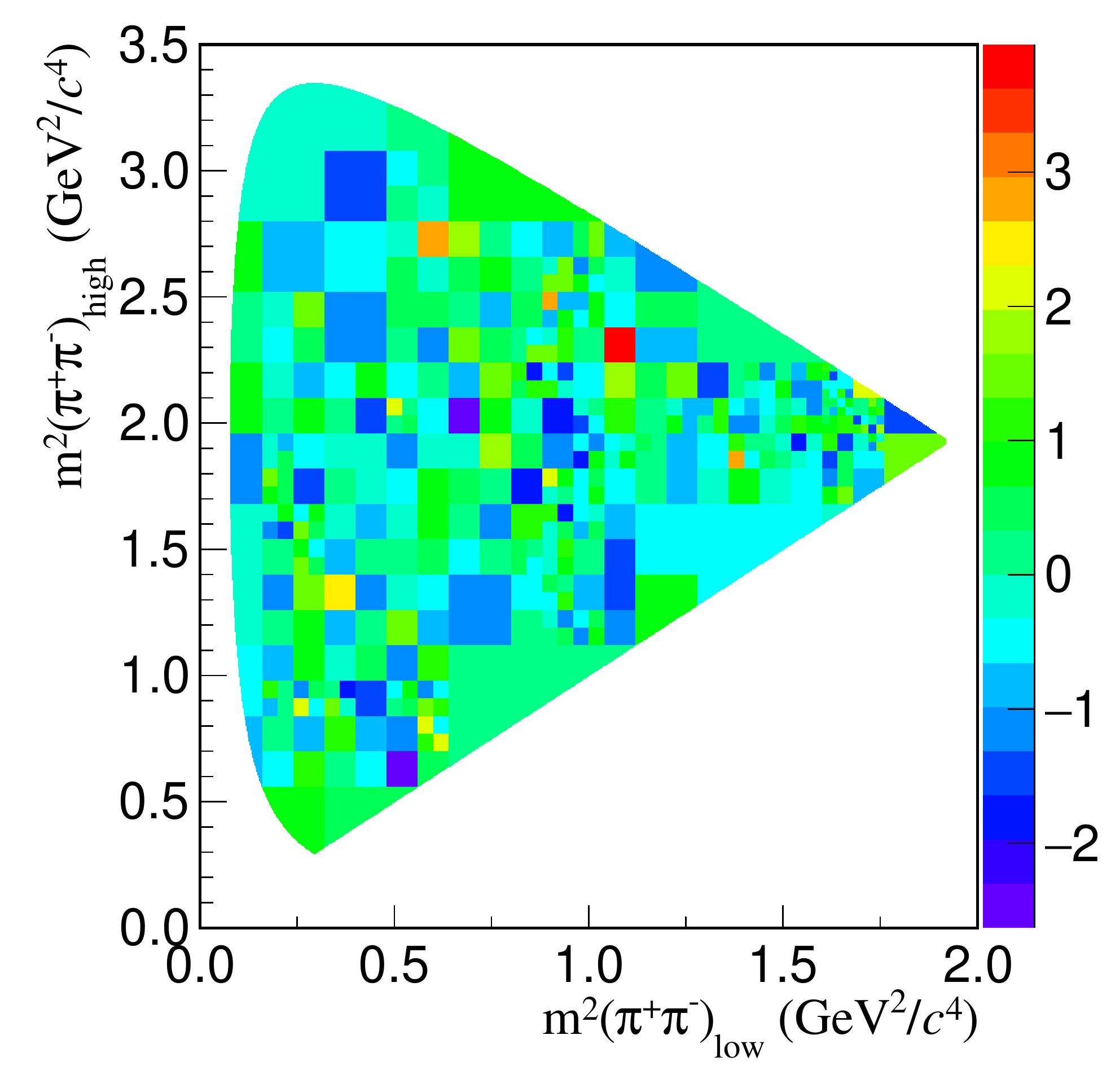}
    \caption{The $\chi$ distribution across the Dalitz plot using an adaptive binning method.}
\label{fig:chisq}
\end{figure}

\subsection{Goodness of fit}\label{sec:PWA_gof}
In order to check the goodness-of-fit of our fit results
quantitatively, we use a two-dimensional $\chi^2$ test by dividing the
Dalitz plot into a number of cells.  For the $i$th cell, we have
$\chi_i = \frac{N_i-N_i^{\rm exp}}{\sqrt{N_i^{\rm exp}}}$, where $N_i$
and $N_i^{\rm exp}$ are the observed number of $\Ds$ candidates and expected
number of $\Ds$ candidates based on the fit model, respectively.  The total
$\chi^2$ by summing up $\chi^2_i$ over all cells divided by the number
of degrees of freedom ($\nu = N_{\rm cell} -N_{\rm par}$, where
$N_{\rm cell}$ is the number of cells having data entries) is used to quantify
the fit quality.  
We calculate $\chi^2$ by
using an adaptive binning process and requiring the minimal number of
entries in each cell is 9, as shown in Fig.~\ref{fig:chisq} for the $\chi_i$
values, 
which leads to $\chi^2/\nu = 344.4/(404-62) $, with a $\chi^2$
probability of $45$\%. Figure~\ref{fig:chisq} also indicates generally good
data and model agreement across the PHSP.

\section{Systematic Uncertainties}\label{sec:uncertainties}
We evaluate systematic uncertainties from the following sources, 
for each source except for the last one the maximum differences between the nominal and varied results 
are taken as the corresponding systematic uncertainty:
\begin{enumerate}[label=\Roman*]
    \item 
        The
      uncertainties arising from the lack of knowledge 
        of the effective barrier radii of mesons are estimated by using alternative values of
        $R_{D_s}$ and $R_{r}$ constants other than the default ones
        [$5\ {(\gevc)}^{-1}$ and $1.5\ {(\gevc)}^{-1}$], within the range 
        $[3.0 - 7.0]\ {(\gevc)}^{-1}$ and $[0.0 - 3.0]\ {(\gevc)}^{-1}$,
      respectively. The variations on $R_{D_s}$ and $R_{r}$ values are done one at a time.
    \item 
        The uncertainties arising from the uncertainties
        on the masses and widths of resonances in the
        baseline model are estimated by varying in turn the masses and widths one standard deviation up and down
        from world averages.
    \item For the uncertainties related to the signal efficiency across the Dalitz plot plane, we vary in turn the
      coefficients used to parametrize Dalitz plot efficiency listed in Table~\ref{tab:effpar} 
        one standard deviation up and down.
    \item For the uncertainties arising from background modeling, instead of the baseline background model, 
        similar to that used in Ref.~\cite{babarpaper}, 
        we use a parametrized background PDF
      by considering contributions from a $\rho(770)$ meson, two
      {\it ad hoc} scalar resonances with free parameters, and a third
        order polynomial.  The contributions are summed incoherently,
      and the background PDF parameters are determined by fitting to the
      sideband data and fixed in the amplitude analysis. In addition, we also model the background contribution using $\Ds$ candidates from only the lower or higher sideband regions.
    \item The uncertainties arising from modeling of $\rho$ resonances are estimated by performing a fit where the
      $\rho$ mesons are parametrized instead by the Gounaris-Sakurai
      formalism~\cite{Gounaris:1968mw}.
    \item As the statistical significance of
      adding the $\omega(782)\pip$ contribution in the baseline signal model
        is below 5$\sigma$, 
        the systematic uncertainties are assigned for those
        parameters that are common between the two models with and without $\omega(782)\pip$.
    \item The uncertainties related to the signal purity are estimated by scaling the signal fraction
        [$F\left(m(\Ds)\right)$ in Eq.~(\ref{eq:LH})] by $\pm1\%$,  
        which is the uncertainty on the signal purity determined from the 
        $m(\Ds)$ fit as depicted in Fig.~\ref{fig:mDsfit}.
    \item The uncertainties related to the $m(\Ds)$ signal region are estimated by fitting 
        to a sample of 12,232 data
      candidates selected within the 
        region of $\vert\!m(\Ds)-m_{D_s}\vert<6$~\mevcc and relaxed
      requirements on the NN responses. The purity is kept at about the
      same level as the nominal sample (80\%).
    \item The uncertainties due to the fit procedure are estimated by generating signal MC candidates using the fitted
      parameters shown in Tables~\ref{tab:fitres}
      and~\ref{tab:fitres_sw}. The signal MC candidates are then mixed
      with candidates from the inclusive MC sample with signal decays removed 
        to form 35 MC
      samples, each with total candidate number and signal purity matched to
      those in data.  The parameters of the 
      control points that are close to the Dalitz plot kinematic
      limits and therefore statistically limited have considerable fit biases, which are about the same
      size as their statistical uncertainties.  We take the mean
      biases observed when fitting to the MC samples as the related
      uncertainties.
\end{enumerate}

Furthermore, concerning the choice of control points, the number of control points used for \swavehy\ modeling has been varied by $\pm 2$ as a consistency check. 
Also, a fit to data is performed with the \swavehy\ amplitude parametrized as an interpolation of magnitudes and phases of the 29 control points, instead of their real and imaginary parts as in the baseline fit.
As no notable variations for the fit parameters are observed in both cases, no systematic uncertainty is assigned. 

Tables~\ref{tab:summ_sys_ff} and~\ref{tab:summ_sys_sw} summarize
contributions from the different systematic sources. These
contributions are combined in quadrature to determine the total
systematic uncertainties.

\begin{table*}
\centering
    \caption{Systematic uncertainties for  fit fractions and
    $\rho$ mesons' coefficients. The dominant systematic uncertainties are highlighted in bold. }
\label{tab:summ_sys_ff}
\begin{tabular}{l|ccccccccc|c}
\hline\hline
    & \RNum{1} & \RNum{2}& \RNum{3}& \RNum{4}& \RNum{5}& \RNum{6}& \RNum{7}& \RNum{8}& \RNum{9} & Total \\
\hline
${\cal F}_{f_2(1270)}$ & ${\bf 0.85}$ & $0.07$ & $0.07$ & $0.35$ & $0.01$ & $0.02$ & $0.12$ & $0.61$ & $0.00$& 1.12\\
${\cal F}_{\rho(770)}$ & $0.10$ & $0.10$ & $0.06$ & $0.10$ & $0.05$ & ${\bf 0.44}$ & $0.09$ & $0.09$ & $0.11$& 0.51\\
${\cal F}_{\rho(1450)}$ & ${\bf 0.42}$ & $0.05$ & $0.05$ & $0.08$ & $0.04$ & $0.16$ & $0.08$ & $0.05$ & $0.14$& 0.49\\
${\cal F}_{{\cal S}{\rm \,wave}}$ & ${\bf 0.67}$ & $0.04$ & $0.10$ & $0.56$ & $0.06$ & $0.02$ & $0.33$ & $0.64$ & $0.44$& 1.21\\\hline
$|c_{\rho(770)}|$ & $0.01$ & $0.00$ & $0.00$ & $0.00$ & $0.01$ & ${\bf 0.04}$ & $0.01$ & $0.00$ & $0.01$& 0.04\\
$\phi_{\rho(770)}$ (rad) & $0.09$ & $0.03$ & $0.04$ & $0.26$ & $0.01$ & ${\bf 0.40}$ & $0.04$ & $0.34$ & $0.05$& 0.60\\
$|c_{\rho(1450)}|$ & $0.08$ & ${\bf 0.11}$ & $0.02$ & $0.04$ & $0.10$ & $0.05$ & $0.03$ & $0.05$ & $0.06$& 0.21\\
$\phi_{\rho(1450)}$ (rad) & $0.09$ & $0.11$ & $0.02$ & $0.05$ & $0.09$ & $0.10$ & $0.10$ & $0.00$ & ${\bf 0.22}$& 0.32\\\hline
\hline
\end{tabular}
\end{table*}

\begin{table*}
\footnotesize
\centering
    \caption{Systematic uncertainties for the parameters of the
    \swavehy\ control points. The dominant systematic uncertainties are highlighted in bold.}
\label{tab:summ_sys_sw}
\begin{tabular}{ll|ccccccccc|c}
\hline
    & & \RNum{1} & \RNum{2}& \RNum{3}& \RNum{4}& \RNum{5}& \RNum{6}& \RNum{7}& \RNum{8}& \RNum{9} & Total\\
\hline
1 & Magnitude  & $0.96$ & $0.07$ & $0.05$ & $0.48$ & $0.03$ & $0.08$ & $0.22$ & $0.80$ & ${\bf 1.07}$& 1.73\\
& Phase (rad) & $0.37$ & $0.13$ & $0.02$ & $0.30$ & $0.10$ & $0.00$ & $0.05$ & $0.42$ & ${\bf 0.98}$& 1.19\\
\hline
2 & Magnitude  & $0.34$ & $0.05$ & $0.05$ & $0.31$ & $0.03$ & $0.05$ & $0.15$ & $0.10$ & ${\bf 0.35}$& 0.62\\
& Phase (rad) & ${\bf 0.17}$ & $0.05$ & $0.01$ & $0.07$ & $0.02$ & $0.03$ & $0.01$ & $0.04$ & $0.05$& 0.21\\
\hline
3 & Magnitude  & $0.18$ & $0.05$ & $0.04$ & $0.31$ & $0.01$ & $0.01$ & $0.16$ & $0.16$ & ${\bf 0.45}$& 0.62\\
& Phase (rad) & ${\bf 0.12}$ & $0.03$ & $0.01$ & $0.04$ & $0.02$ & $0.02$ & $0.00$ & $0.00$ & $0.04$& 0.14\\
\hline
4 & Magnitude  & $0.11$ & $0.02$ & $0.04$ & ${\bf 0.30}$ & $0.03$ & $0.14$ & $0.15$ & $0.18$ & $0.02$& 0.42\\
& Phase (rad) & ${\bf 0.07}$ & $0.03$ & $0.01$ & $0.05$ & $0.02$ & $0.06$ & $0.00$ & $0.05$ & $0.04$& 0.13\\
\hline
5 & Magnitude  & $0.26$ & $0.03$ & $0.05$ & ${\bf 0.46}$ & $0.06$ & $0.02$ & $0.17$ & $0.11$ & $0.19$& 0.60\\
& Phase (rad) & $0.04$ & $0.01$ & $0.01$ & $0.01$ & $0.02$ & $0.02$ & $0.02$ & $0.01$ & ${\bf 0.07}$& 0.09\\
\hline
6 & Magnitude  & ${\bf 0.43}$ & $0.06$ & $0.06$ & $0.27$ & $0.02$ & $0.03$ & $0.18$ & $0.02$ & $0.26$& 0.61\\
& Phase (rad) & $0.03$ & $0.01$ & $0.01$ & $0.03$ & $0.00$ & $0.04$ & $0.01$ & $0.02$ & ${\bf 0.12}$& 0.14\\
\hline
7 & Magnitude  & ${\bf 0.54}$ & $0.07$ & $0.04$ & $0.22$ & $0.00$ & $0.26$ & $0.13$ & $0.03$ & $0.06$& 0.66\\
& Phase (rad) & $0.06$ & $0.01$ & $0.01$ & $0.02$ & $0.01$ & $0.00$ & $0.02$ & ${\bf 0.06}$ & $0.03$& 0.09\\
\hline
8 & Magnitude  & ${\bf 0.60}$ & $0.12$ & $0.06$ & $0.25$ & $0.00$ & $0.01$ & $0.11$ & $0.59$ & $0.09$& 0.90\\
& Phase (rad) & $0.06$ & $0.02$ & $0.00$ & $0.01$ & $0.01$ & $0.02$ & $0.01$ & $0.02$ & ${\bf 0.07}$& 0.10\\
\hline
9 & Magnitude  & ${\bf 0.65}$ & $0.17$ & $0.08$ & $0.25$ & $0.01$ & $0.01$ & $0.08$ & $0.56$ & $0.14$& 0.93\\
& Phase (rad) & ${\bf 0.05}$ & $0.02$ & $0.01$ & $0.01$ & $0.02$ & $0.04$ & $0.01$ & $0.03$ & $0.03$& 0.08\\
\hline
10 & Magnitude  & $0.69$ & $0.21$ & $0.10$ & $0.27$ & $0.01$ & $0.06$ & $0.07$ & ${\bf 1.15}$ & $0.21$& 1.41\\
& Phase (rad) & $0.05$ & $0.02$ & $0.01$ & $0.02$ & $0.02$ & $0.03$ & $0.01$ & ${\bf 0.06}$ & $0.01$& 0.09\\
\hline
11 & Magnitude  & $0.72$ & $0.25$ & $0.11$ & $0.25$ & $0.00$ & $0.09$ & $0.07$ & ${\bf 0.87}$ & $0.74$& 1.40\\
& Phase (rad) & ${\bf 0.05}$ & $0.02$ & $0.01$ & $0.02$ & $0.02$ & $0.04$ & $0.01$ & $0.01$ & $0.03$& 0.07\\
\hline
12 & Magnitude  & $0.43$ & $0.18$ & $0.07$ & $0.19$ & $0.00$ & $0.06$ & $0.06$ & ${\bf 0.83}$ & $0.03$& 0.98\\
& Phase (rad) & ${\bf 0.05}$ & $0.02$ & $0.01$ & $0.02$ & $0.02$ & $0.04$ & $0.01$ & $0.01$ & $0.03$& 0.08\\
\hline
13 & Magnitude  & $0.15$ & $0.09$ & $0.04$ & $0.19$ & $0.01$ & $0.04$ & $0.07$ & ${\bf 0.35}$ & $0.03$& 0.44\\
& Phase (rad) & ${\bf 0.05}$ & $0.01$ & $0.00$ & $0.02$ & $0.02$ & $0.03$ & $0.00$ & $0.01$ & $0.04$& 0.07\\
\hline
14 & Magnitude  & $0.06$ & $0.06$ & $0.02$ & ${\bf 0.13}$ & $0.02$ & $0.03$ & $0.04$ & $0.04$ & $0.03$& 0.17\\
& Phase (rad) & $0.02$ & $0.01$ & $0.00$ & $0.03$ & $0.01$ & ${\bf 0.03}$ & $0.01$ & $0.02$ & $0.01$& 0.05\\
\hline
15 & Magnitude  & $0.13$ & $0.05$ & $0.02$ & $0.05$ & $0.03$ & $0.06$ & $0.02$ & ${\bf 0.14}$ & $0.02$& 0.22\\
& Phase (rad) & $0.01$ & $0.01$ & $0.00$ & $0.04$ & $0.01$ & $0.02$ & $0.01$ & ${\bf 0.05}$ & $0.00$& 0.07\\
\hline
16 & Magnitude  & $0.12$ & $0.05$ & $0.02$ & $0.07$ & $0.02$ & $0.04$ & $0.03$ & ${\bf 0.42}$ & $0.10$& 0.46\\
& Phase (rad) & $0.01$ & $0.01$ & $0.00$ & ${\bf 0.01}$ & $0.01$ & $0.01$ & $0.00$ & $0.00$ & $0.01$& 0.03\\
\hline
17 & Magnitude  & $0.17$ & $0.06$ & $0.02$ & $0.07$ & $0.01$ & $0.04$ & $0.03$ & ${\bf 0.48}$ & $0.08$& 0.53\\
& Phase (rad) & ${\bf 0.02}$ & $0.01$ & $0.00$ & $0.02$ & $0.01$ & $0.01$ & $0.01$ & $0.00$ & $0.01$& 0.03\\
\hline
18 & Magnitude  & $0.20$ & $0.05$ & $0.02$ & $0.08$ & $0.00$ & $0.02$ & $0.03$ & ${\bf 0.36}$ & $0.03$& 0.42\\
& Phase (rad) & $0.02$ & $0.01$ & $0.00$ & $0.01$ & $0.00$ & $0.01$ & $0.00$ & ${\bf 0.08}$ & $0.03$& 0.09\\
\hline
19 & Magnitude  & ${\bf 0.21}$ & $0.05$ & $0.02$ & $0.08$ & $0.02$ & $0.02$ & $0.04$ & $0.06$ & $0.03$& 0.25\\
& Phase (rad) & $0.03$ & $0.01$ & $0.00$ & $0.01$ & $0.00$ & $0.01$ & $0.00$ & $0.01$ & ${\bf 0.05}$& 0.06\\
\hline
20 & Magnitude  & ${\bf 0.28}$ & $0.06$ & $0.03$ & $0.08$ & $0.03$ & $0.00$ & $0.02$ & $0.20$ & $0.10$& 0.38\\
& Phase (rad) & ${\bf 0.04}$ & $0.01$ & $0.00$ & $0.01$ & $0.00$ & $0.00$ & $0.00$ & $0.03$ & $0.03$& 0.06\\
\hline
21 & Magnitude  & $0.37$ & $0.06$ & $0.03$ & $0.04$ & $0.02$ & $0.02$ & $0.01$ & ${\bf 0.57}$ & $0.05$& 0.69\\
& Phase (rad) & $0.05$ & $0.01$ & $0.00$ & $0.02$ & $0.00$ & $0.01$ & $0.01$ & ${\bf 0.07}$ & $0.05$& 0.11\\
\hline
22 & Magnitude  & ${\bf 0.40}$ & $0.07$ & $0.03$ & $0.03$ & $0.02$ & $0.02$ & $0.00$ & $0.15$ & $0.12$& 0.45\\
& Phase (rad) & $0.05$ & $0.02$ & $0.00$ & $0.02$ & $0.00$ & $0.01$ & $0.01$ & ${\bf 0.09}$ & $0.06$& 0.12\\
\hline
23 & Magnitude  & ${\bf 0.43}$ & $0.09$ & $0.03$ & $0.06$ & $0.00$ & $0.03$ & $0.01$ & $0.29$ & $0.43$& 0.68\\
& Phase (rad) & ${\bf 0.06}$ & $0.02$ & $0.00$ & $0.01$ & $0.00$ & $0.02$ & $0.01$ & $0.02$ & $0.01$& 0.07\\
\hline
24 & Magnitude  & ${\bf 0.44}$ & $0.09$ & $0.03$ & $0.10$ & $0.01$ & $0.02$ & $0.04$ & $0.28$ & $0.16$& 0.56\\
& Phase (rad) & ${\bf 0.07}$ & $0.02$ & $0.00$ & $0.01$ & $0.01$ & $0.03$ & $0.02$ & $0.04$ & $0.03$& 0.09\\
\hline
25 & Magnitude  & ${\bf 0.42}$ & $0.10$ & $0.04$ & $0.14$ & $0.02$ & $0.04$ & $0.04$ & $0.26$ & $0.35$& 0.63\\
& Phase (rad) & ${\bf 0.07}$ & $0.02$ & $0.00$ & $0.01$ & $0.01$ & $0.03$ & $0.02$ & $0.02$ & $0.02$& 0.09\\
\hline
26 & Magnitude  & $0.30$ & $0.09$ & $0.04$ & $0.15$ & $0.00$ & $0.10$ & $0.06$ & ${\bf 0.41}$ & $0.27$& 0.61\\
& Phase (rad) & ${\bf 0.07}$ & $0.04$ & $0.00$ & $0.02$ & $0.02$ & $0.04$ & $0.02$ & $0.07$ & $0.02$& 0.12\\
\hline
27 & Magnitude  & $0.28$ & $0.19$ & $0.08$ & $0.17$ & $0.04$ & $0.17$ & $0.05$ & ${\bf 0.64}$ & $0.17$& 0.79\\
& Phase (rad) & ${\bf 0.32}$ & $0.17$ & $0.01$ & $0.23$ & $0.03$ & $0.10$ & $0.23$ & $0.06$ & $0.23$& 0.56\\
\hline
28 & Magnitude  & ${\bf 0.78}$ & $0.19$ & $0.05$ & $0.23$ & $0.12$ & $0.13$ & $0.14$ & $0.08$ & $0.60$& 1.06\\
& Phase (rad) & ${\bf 0.53}$ & $0.09$ & $0.06$ & $0.06$ & $0.03$ & $0.07$ & $0.17$ & $0.48$ & $0.21$& 0.78\\
\hline
29 & Magnitude  & $0.43$ & $0.09$ & $0.05$ & $0.15$ & $0.05$ & $0.42$ & $0.24$ & $0.34$ & ${\bf 1.52}$& 1.70\\
& Phase (rad) & ${\bf 0.55}$ & $0.03$ & $0.01$ & $0.04$ & $0.05$ & $0.21$ & $0.06$ & $0.09$ & $0.07$& 0.60\\
\hline
\hline
\end{tabular}
\end{table*}
\section{Conclusion}
\label{sec:CONLUSION}

Based on 3.19 fb$^{-1}$ of data taken at $E_{\rm c.m.}=4.178$ GeV
with the BESIII detector at the BEPCII collider, we select a sample of
13,797 $\Ds\to\pipipi$ candidates with a signal purity of 80\%. 
The amplitude analysis shows the decay is dominated by the $\pipi$ \swave.
We also observe a significant spin-2 contribution with the 
fit fraction consistent with that reported by BABAR.
Our fit fraction result 
of ${\cal F}(\Ds\to \rho^0\pip) 
= (0.9\pm 0.4 _{\rm stat}\pm 0.5_{\rm syst})\%$ shows a central value somewhat lower than that of the BABAR\ result, 
however the two results are still compatible within one standard deviation. 
Based on the known ${\cal B}(\Ds\to\pip\pim\pip)$~\cite{pdg},
we have ${\cal B}(\Ds\to\rho^0\pip) = {\cal B}(\Ds\to\pip\pim\pip)\times {\cal F}(\Ds\to \rho^0\pip)  = (0.009 \pm 0.007)$\% that agrees with the predictions 
in Ref.~\cite{Li:2013xsa}.


As using
 relativistic Breit-Wigner PDFs to model overlapping intermediate
scalars such as $f_0(980)$
 and $f_0(1370)$ will lead to a violation of unitarity and is thus unphysical, the \swavehy\ content is determined using a
 quasi-model-independent partial-wave-analysis method.
 Our results show good agreement with BABAR\ with a similar data sample size~\cite{babarpaper}. The statistical uncertainties of our results
 are generally better than the BABAR\ ones. 
 As the same choice of control points on the $m(\pipi)$ spectrum is used, 
 combining \swavehy\ results from both BESIII and BABAR\ could offer
 a very precise description of the $\pipi$ \swave\ in $\Ds\to\pipipi$, which can be later used to test new models for light scalar resonances. 

\begin{acknowledgements}
\label{sec:acknowledgement}
\vspace{-0.4cm}
The BESIII collaboration thanks the staff of BEPCII and the IHEP computing center for their strong support. This work is supported in part by National Key R\&D Program of China under Contracts Nos. 2020YFA0406400, 2020YFA0406300; National Natural Science Foundation of China (NSFC) under Contracts Nos. 11605124, 11625523, 11635010, 11735014, 11822506, 11835012, 11935015, 11935016, 11935018, 11961141012, 12022510, 12025502, 12035009, 12035013, 12061131003; the Chinese Academy of Sciences (CAS) Large-Scale Scientific Facility Program; Joint Large-Scale Scientific Facility Funds of the NSFC and CAS under Contracts Nos. U1732263, U1832207, U1932108; CAS Key Research Program of Frontier Sciences under Contract No. QYZDJ-SSW-SLH040; 100 Talents Program of CAS; INPAC and Shanghai Key Laboratory for Particle Physics and Cosmology; ERC under Contract No. 758462; European Union Horizon 2020 research and innovation programme under Contract No. Marie Sklodowska-Curie grant agreement No 894790; German Research Foundation DFG under Contracts Nos. 443159800, Collaborative Research Center CRC 1044, FOR 2359, FOR 2359, GRK 214; Istituto Nazionale di Fisica Nucleare, Italy; Ministry of Development of Turkey under Contract No. DPT2006K-120470; National Science and Technology fund; Olle Engkvist Foundation under Contract No. 200-0605; STFC (United Kingdom); The Knut and Alice Wallenberg Foundation (Sweden) under Contract No. 2016.0157; The Royal Society, UK under Contracts Nos. DH140054, DH160214; The Swedish Research Council; U. S. Department of Energy under Contracts Nos. DE-FG02-05ER41374, DE-SC-0012069.
\end{acknowledgements}

\clearpage
\appendix
\section{Correlation matrices for the fit fractions}
\label{sec:appB}

The statistical and systematic correlation matrices for the fit fractions are 
given in Tables~\ref{tab:statcorrmatff} and~\ref{tab:systcorrmatff}. 
The statistical and systematic correlation matrices for the 62 fit parameters
shown in Table~\ref{tab:fitres} and Table~\ref{tab:fitres_sw} are given in 
the Supplemental Material~\cite{supp}. 

\begin{table}[!htb]
\centering
    \caption{Statistical correlation matrix for the fit fractions. }
\label{tab:statcorrmatff}
\begin{tabular}{ccccc}
\hline\hline
& $f_2(1270)\pip$& $\rho(770)\pip$& $\rho(1450)\pip$& \swave
\\\hline
$f_2(1270)\pip$ & $+1.00$ & $-0.15$ & $-0.30$ & $-0.10$\\
$\rho(770)\pip$ &  & $+1.00$ & $-0.40$ & $-0.50$\\
$\rho(1450)\pip$ &  &  & $+1.00$ & $+0.27$\\
\swave &  &  &  & $+1.00$\\
\hline
\hline
\end{tabular}
\end{table}

\begin{table}[!htb]
\centering
    \caption{Systematic correlation matrix for the fit fractions. }
\label{tab:systcorrmatff}
\begin{tabular}{ccccc}
\hline\hline
& $f_2(1270)\pip$& $\rho(770)\pip$& $\rho(1450)\pip$& \swave
\\\hline
$f_2(1270)\pip$ &  $+1.00$  &  $+0.31$  &  $+0.50$  &  $-0.04$ \\
$\rho(770)\pip$ &   &  $+1.00$  &  $-0.16$  &  $-0.06$ \\
$\rho(1450)\pip$ &   &   &  $+1.00$  &  $+0.74$ \\
\swave &   &   &   &  $+1.00$ \\
\hline
\hline
\end{tabular}
\end{table}


\end{document}